\documentclass{cimento}
\usepackage{amsmath}
\usepackage{tabularx}
\usepackage{graphicx}  
\usepackage{caption}
\usepackage{subcaption}

\title{Recent results from T2K}
\author{M.~Scott,\thanks{On behalf of the T2K collaboration}}
\instlist{\inst{TRIUMF}, Canada}

\PACSes{\PACSit{14.60}\PACSit{13.15}}

\begin{document}

\maketitle

\begin{abstract}
The T2K long-baseline neutrino oscillation experiment has produced the first observation of $\nu_{e}$ appearance and the most precise measurement of the mixing angle $\theta_{23}$ from $6.57 \times 10^{20}$ protons-on-target (POT) of neutrino beam data.
We present here the results of T2K antineutrino oscillation analyses searching for $\bar{\nu}_{e}$ appearance and measuring the dominant oscillation parameters for $\bar{\nu}_{\mu}$ disappearance, including the results from the T2K near detector fit.
Using $4.01 \times 10^{20}$ POT of antineutrino beam data T2K measured $\sin^{2}(\bar{\theta}_{23}) = 0.45^{+0.29}_{-0.12}$ and $|\Delta \bar{m}^{2}_{32}| = 2.51\pm0.28 \times 10^{-3} (eV^{2})$, consistent with previous existing $\nu_{\mu}$ and $\bar{\nu}_{\mu}$ disappearance measurements.
From the same dataset 3 $\bar{\nu}_{e}$ candidate events were selected at the far detector, which did not provide evidence either for or against the $\bar{\nu}_{e}$ appearance hypothesis.
\end{abstract}

\section{Introduction}
The discovery of neutrino oscillations, and the implication that neutrinos are massive particles, is the only conclusive evidence to date for new physics.
Explaining neutrino masses is a necessary part of any theory aiming to go beyond the Standard Model and so studying neutrinos and neutrino oscillations is a powerful method of constraining or eliminating these theories.

In the Pontecorvo-Maki-Nakagawa-Sakata (PMNS) formalism, neutrino oscillation is parameterised by three mixing angles, $\theta_{12}$, $\theta_{23}$ and $\theta_{13}$, two mass squared splittings, $\Delta m^{2}_{21}$ and $\Delta m^{2}_{32}$, and one CP violating phase, $\delta_{\textrm{CP}}$.  
These parameters are encapsulated in a mixing matrix, $U$, which relates the neutrino weak flavour eigenstates to the neutrino mass eigenstates, shown in Eq.~(\ref{e.u}), where $c_{ij} = \cos\theta_{ij}$ and $s_{ij} = \sin\theta_{ij}$:
\begin{equation}
    \label{e.u}
    U = \left( \begin{array}{ccc}
    1 & 0 & 0 \\
    0 & +c_{23} & +s_{23} \\
    0 & -s_{23} & +c_{23} \end{array} \right) 
    \left( \begin{array}{ccc}
    +c_{13} & 0 & +s_{13}e^{-i\delta} \\ 
    0 & 1 & 0 \\
    -s_{13}e^{+i\delta} & 0 & +c_{13} \end{array} \right) 
    \left( \begin{array}{ccc}
    +c_{12} & +s_{12} & 0 \\
    -s_{12} & +c_{12} & 0 \\
    0 & 0 & 1 \end{array} \right) 
\end{equation}
The probability for a neutrino to be created as one flavour, $\alpha$, and then observed some distance away, $L$, as flavour $\beta$ is then given by Eq.~(\ref{e.osc}):
\begin{equation}
    \label{e.osc}
    P_{\alpha \to \beta} = \left| \displaystyle\sum_{i} U^{*}_{\alpha i}U_{\beta i}~e^{-im^{2}_{i} \frac{L}{2E_{\nu}}}\right|^{2}
\end{equation}

There are currently three open questions in the PMNS oscillation framework:
\begin{itemize}
    \item What is the sign of $\Delta m^{2}_{32}$, or do the neutrino mass states follow the ``normal'' ($\Delta m^{2}_{32} > 0$) or the ``inverse'' ($\Delta m^{2}_{32} < 0$) hierarchy?
    \item Is $\theta_{23}$ greater than $45^{\circ}$?
    \item Is $\sin \delta_{\textrm{CP}} \neq 0$?
\end{itemize}

\begin{figure}[htbp]
    \centering
    \begin{subfigure}[t]{0.31\textwidth}
        \includegraphics[width=\textwidth]{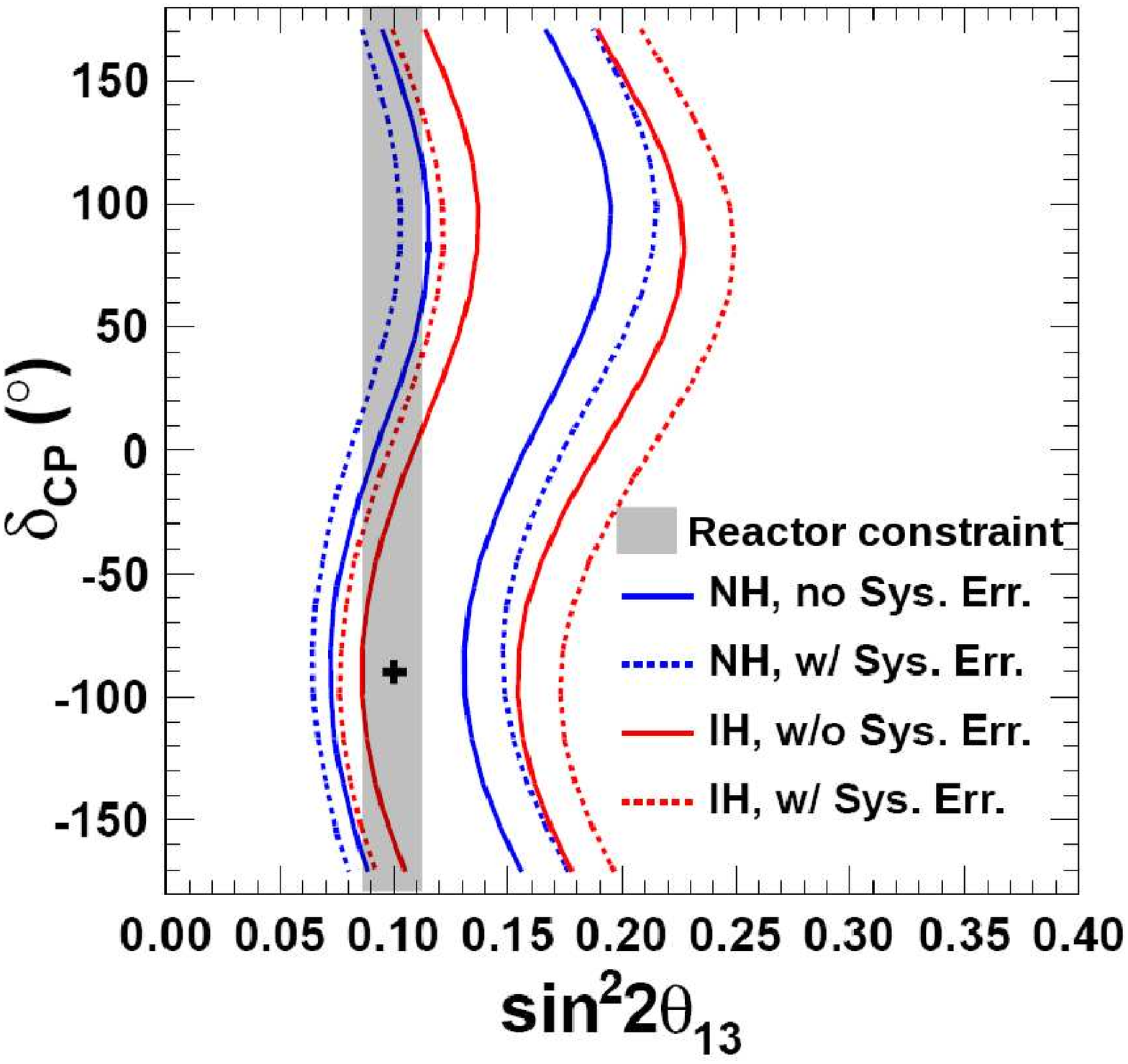}
        \caption{Fitting neutrino data alone}
        \label{fig:nu_only}
    \end{subfigure}
    \hspace{0.01\textwidth}
    \begin{subfigure}[t]{0.31\textwidth}
        \includegraphics[width=\textwidth]{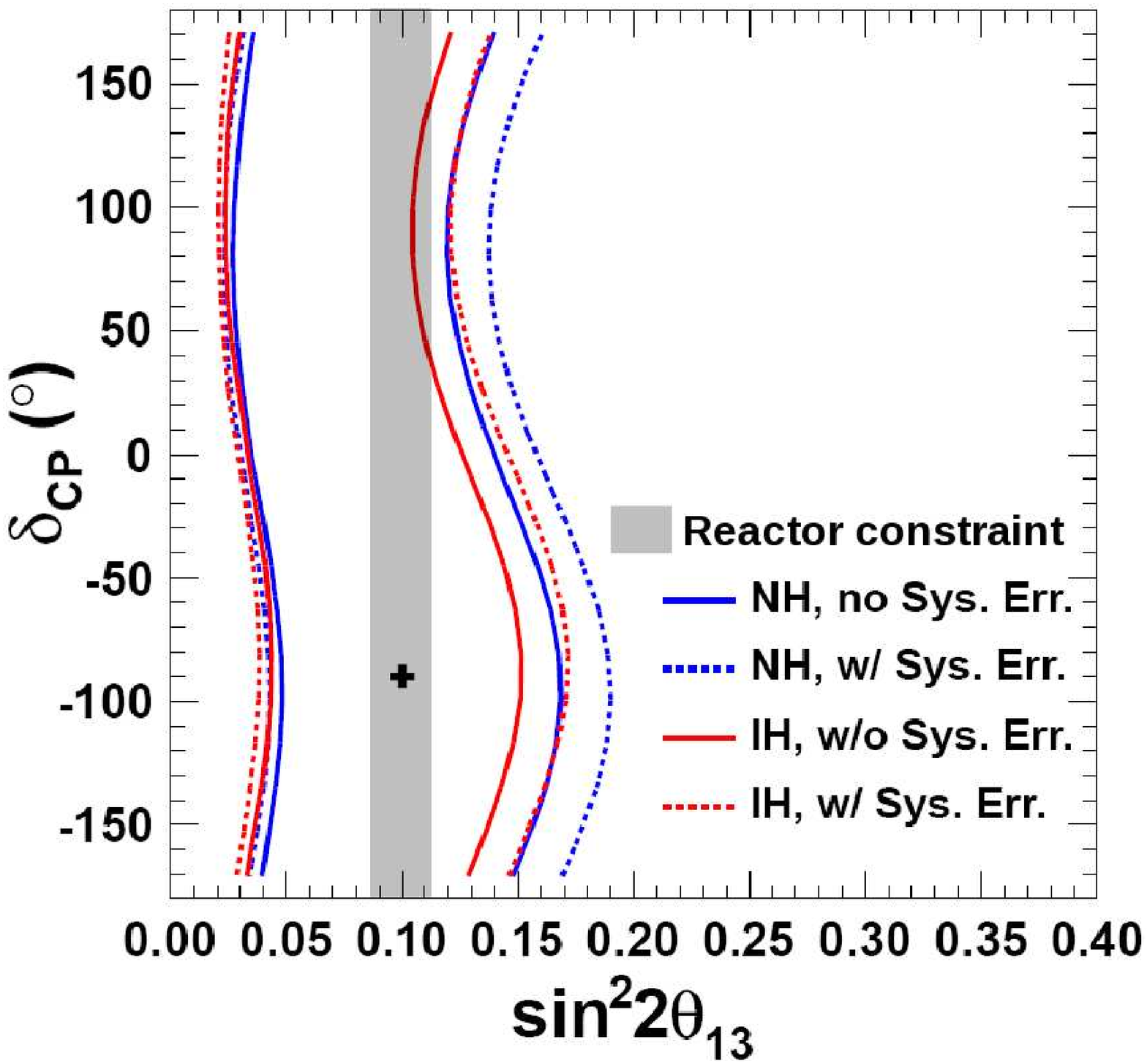}
        \caption{Fitting anti-neutrino data alone}
        \label{fig:anu_only}
    \end{subfigure}
    \hspace{0.01\textwidth}
    \begin{subfigure}[t]{0.305\textwidth}
        \includegraphics[width=\textwidth]{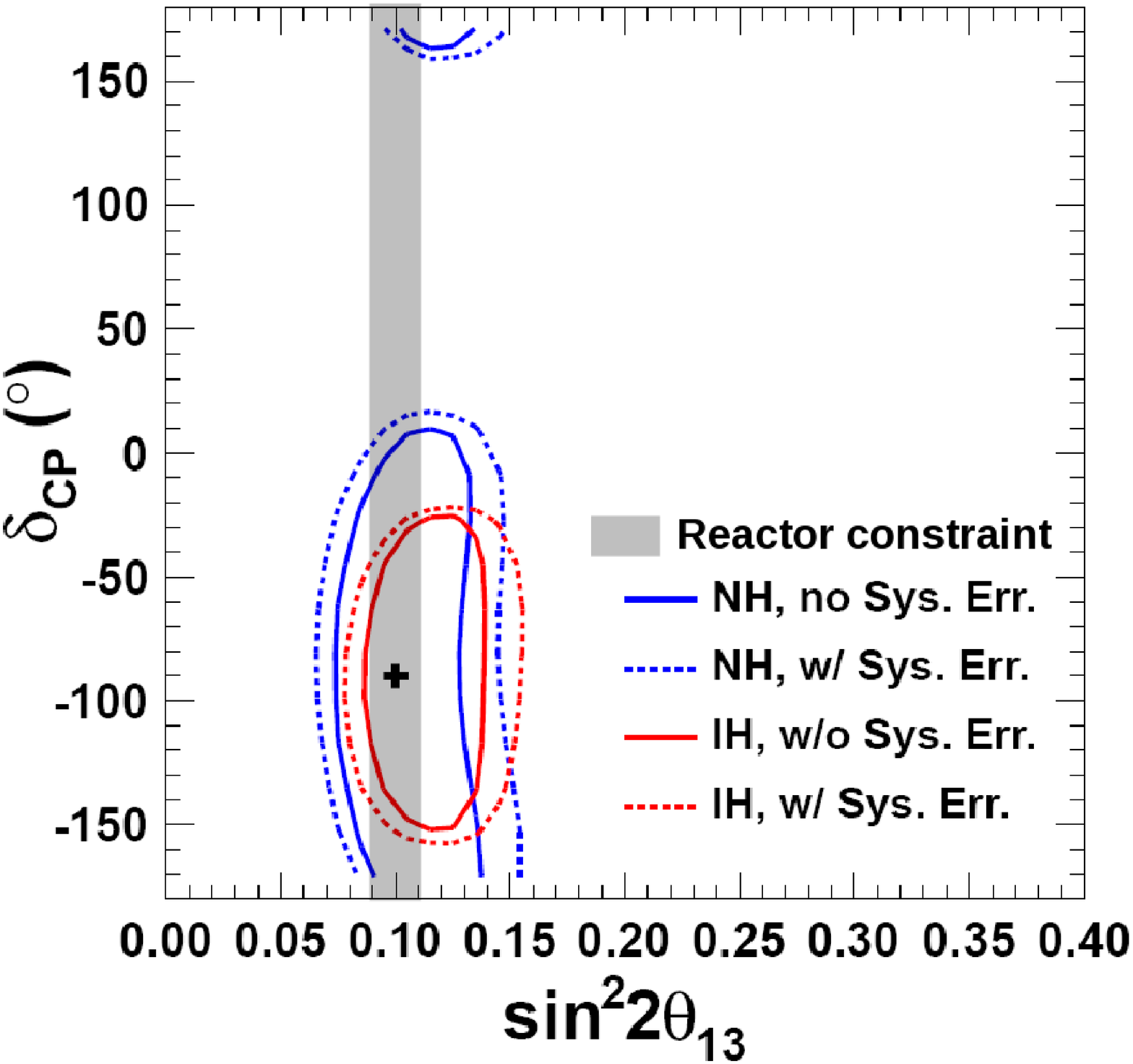}
        \caption{Fitting both neutrino and anti-neutrino data simultaneously}
        \label{fig:both}
    \end{subfigure}
    \caption{Expected 90\% confidence limits on $\delta_{\textrm{CP}}$ and $\sin^{2}\theta_{13}$ from fitting the ultimate T2K dataset assuming a 50:50 split in integrated POT between $\nu$ and $\bar{\nu}$ beam modes~\cite{ref:future}.  The results are for a true value of $\delta_{\textrm{CP}} = -\pi/2$, $\sin^{2}\theta_{13} = 0.1$ and $\Delta m^{2}_{32} > 0$ (the ``Normal'' mass hierarchy, NH).  The solid lines show the fit results with statistical errors only, the dashed lines include systematic uncertainties as well.  The blue lines show fit results assuming NH while the red lines assume the inverted hierarchy, $\Delta m^{2}_{32} < 0$.  The grey band has been added to the figures to show the systematic uncertainty on $\sin^{2}\theta_{13}$ from the Daya Bay experiment~\cite{ref:daya}.}
    \label{fig:anti_nu_osc}
\end{figure}

The Tokai-to-Kamiokande (T2K) experiment has begun answering these questions, with the first observation of $\nu_{e}$ appearance~\cite{ref:nue}, the world's best measurement of $\sin^{2}\theta_{23}$~\cite{ref:joint} and a constraint on $\delta_{\textrm{CP}}$~\cite{ref:joint}.  
These measurements were made using the T2K neutrino beam mode data, where positive pions are focussed to produce a beam mainly composed of $\nu_{\mu}$s.
Since then T2K has been collecting data in anti-neutrino beam mode, with negative pions focussed to give a $\bar{\nu}_{\mu}$ beam.
Measuring both $\nu_{e}$ and $\bar{\nu}_{e}$ appearance breaks the degeneracy between $\delta_{\textrm{CP}}$ and $\sin^{2}\theta_{13}$, as shown in Fig.~\ref{fig:anti_nu_osc}, allowing a constraint to be placed on $\delta_{\textrm{CP}}$ using data from a single experiment.
Comparing these results with measurements of $\sin^{2}2\theta_{13}$ from reactor experiments provides a test of the PMNS oscillation framework, while $\bar{\nu}_{\mu}$ disappearance measurements can also be used to test the CPT symmetry and to search for non-standard interactions of neutrinos with matter.

\section{The T2K experiment}
T2K is a long-baseline neutrino oscillation experiment in Japan, with the neutrino beamline and near detector complex located in Tokai Village, on the east coast, and the far detector, Super-Kamiokande (SK) situated 295~km away in Kamioka~\cite{ref:nim}.

\subsection{Neutrino beamline and flux}
The T2K neutrino beam is a tertiary beam created by colliding protons with a stationary graphite target.
This produces a large number of charged hadrons, largely composed of pions and kaons, that are focussed by three magnetic horns.
The hadrons then enter a 96~m volume where they decay in flight to produce neutrinos.

The T2K detectors are placed $2.5^{\circ}$ off the neutrino beam axis.
This so called `off-axis' method produces a neutrino flux peaked at a lower energy than the on-axis flux and with many fewer neutrinos with energies above 1~GeV.

The results presented here use three data sets: two where the beam operated in anti-neutrino mode, (1) June 2014 and (2) November 2014 -- June 2015, and one in neutrino mode, (3) November 2010 -- May 2013.
The oscillation analyses use data sets (1) and (2), whilst the near detector analysis uses data sets (1) and (3).
This gives an exposure of $4.01 \times 10^{20}$ protons-on-target (POT) in anti-neutrino mode for the oscillation analysis, and an exposure of $0.43 \times 10^{20}$ POT in anti-neutrino mode plus $5.82 \times 10^{20}$ POT in neutrino mode for the near detector analysis.

In anti-neutrino mode, though the flux is dominated by $\bar{\nu}_{\mu}$ there is a significant contamination of $\nu_{\mu}$ events in the SK sample, due to the larger $\nu_{\mu}$ interaction cross section.
Since SK cannot differentiate between neutrino and anti-neutrino interactions this background must be constrained using the T2K near detectors.

\subsection{INGRID and ND280 - the T2K near detectors}
\begin{figure}[htbp]
    \centering
    \begin{subfigure}[t]{0.4\textwidth}
        \includegraphics[width=\textwidth]{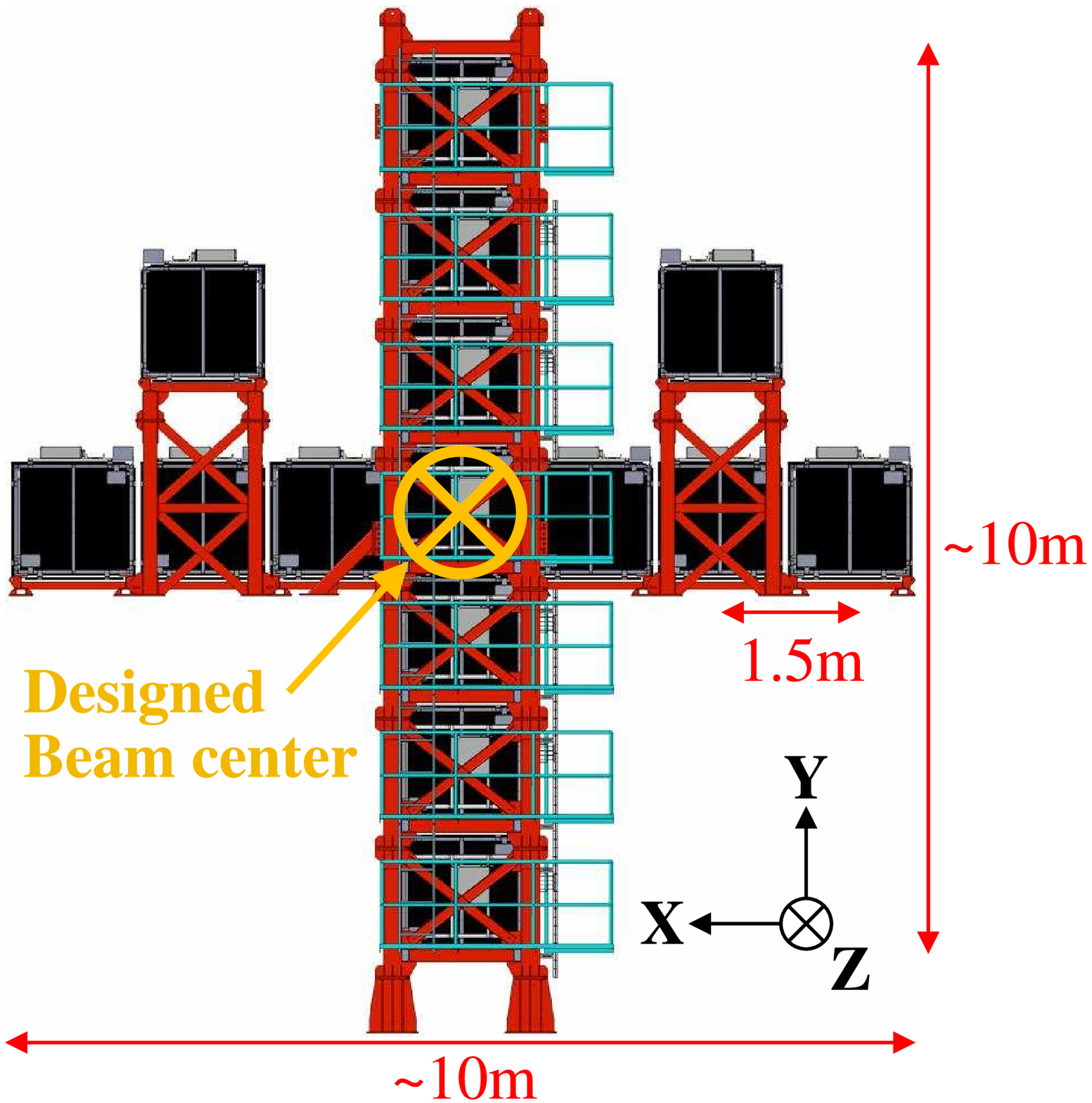}
        \caption{INGRID}
        \label{fig:ingrid}
    \end{subfigure}
    \hspace{0.001\textwidth}
    \begin{subfigure}[t]{0.4\textwidth}
        \includegraphics[width=\textwidth]{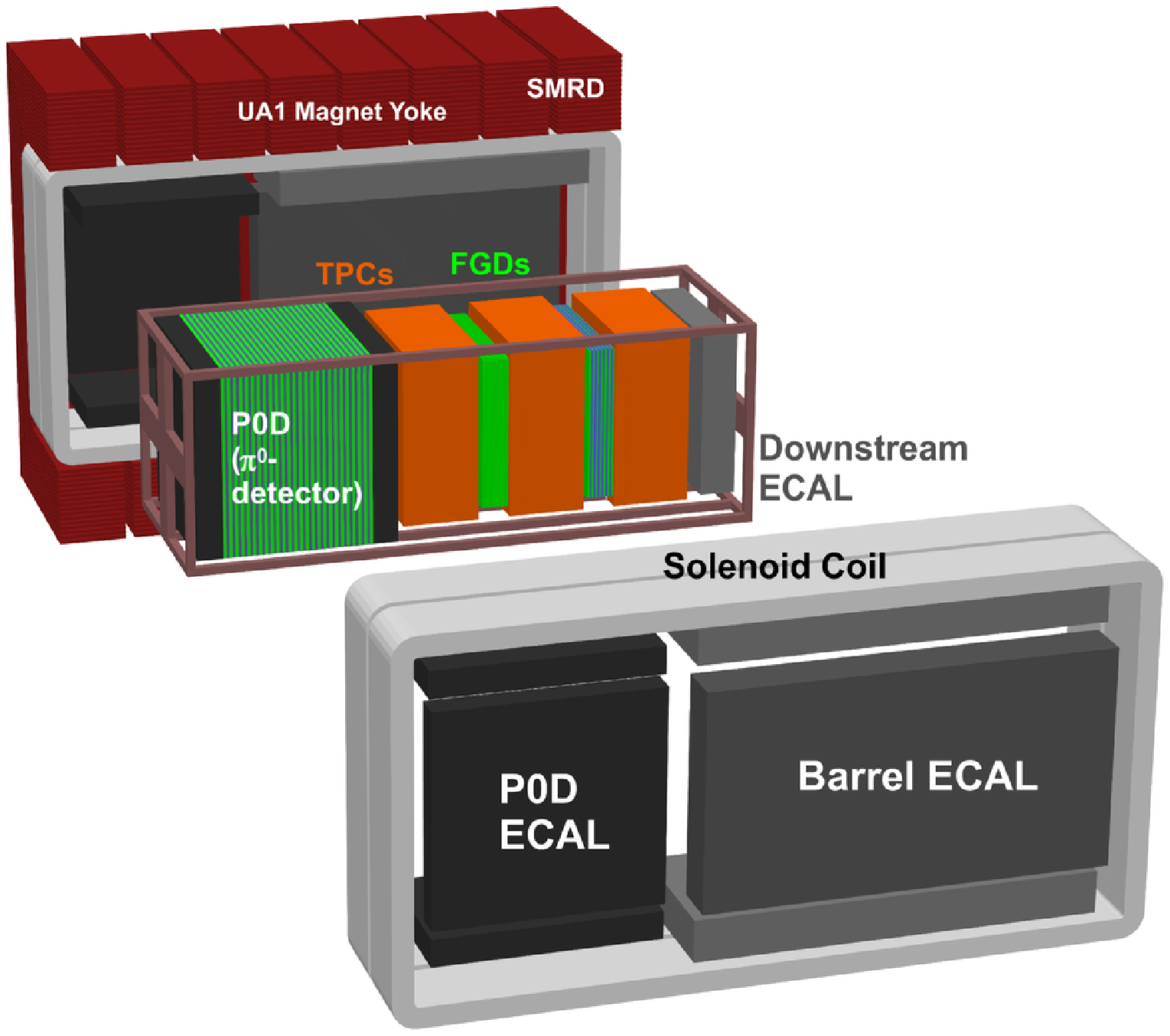}
        \caption{ND280}
        \label{fig:nd280}
        \end{subfigure}
    \caption{The T2K near detectors}
    \label{fig:near_det}
\end{figure}

The near detector facility is positioned 280~m from the hadron production target and contains an on-axis detector, the Interactive Neutrino GRID (INGRID), and an off-axis detector, ND280.
INGRID~\cite{ref:ingrid}, shown in Fig.~\ref{fig:ingrid}, is formed from 14 modules, each of which is composed of layers of iron and plastic scintillator.
This provides a large target mass and therefore a high neutrino interaction rate, allowing INGRID to measure the neutrino beam direction on a spill-by-spill basis.

ND280, shown in Fig.~\ref{fig:nd280}, is a more complex detector which was designed to characterise the neutrino beam observed by SK before any oscillation had occurred.
It has two targets, fine-grained detectors (FGD) 1 and 2, which separate three time projection chambers (TPC), all of which is surrounded by calorimeters and the former UA1/NOMAD magnet, which generates a 0.2~T field perpendicular to the neutrino beam direction.
FGD1 is a fully active carbon target, formed from plastic scintillator bars, while FGD2 also contains a passive water region.
Neutrinos interact in an FGD, which records any short tracks from the interaction, while the TPCs provide particle identification and momentum measurements for particles that exit the FGD.

\subsection{Super-Kamiokande - the T2K far detector}
Super-Kamiokande~\cite{ref:sknim} is 50~kt water Cherenkov detector, with $\sim11000$ 20'' photo-multiplier tubes (PMTs) instrumenting the inner detector and $\sim2000$ 8'' PMTs in the outer detector.
SK uses the shape of the observed Cherenkov ring to differentiate between muons and electrons, and so select muon or electron neutrino interactions.
The event selection requires that the reconstructed interaction vertex is at least 2~m from the inner detector wall, giving a fiducial mass of 22.5~kt.

\section{The T2K oscillation analysis method}
The T2K analysis extrapolates data from the near detectors to predict the far detector event rate using parameterised flux, neutrino cross section and detector models.
The flux and cross section models are initially constrained using data from other experiments, along with data from INGRID and the beamline monitors.
A two-stage fit is then performed, initially to the ND280 data to produce a tuned model that predicts the SK event rate in the absence of oscillation.
This model is then fit to SK data to extract the neutrino oscillation parameters.

\subsection{ND280 data and analysis}
The ND280 event selection chooses the highest momentum, muon-like track as the muon candidate, with the track charge used to differentiate between neutrino and anti-neutrino interactions when using anti-neutrino beam mode data.
In neutrino beam mode the selected events are further separated according to the number of pions observed in the detector, giving charged current $0\pi$, $1\pi$ and $N\pi$ samples.
In anti-neutrino beam mode the events are split according to the total track multiplicity into separate 1-Track and N-Track samples for both positive and negative muon candidates.

Experimentalists cannot measure the neutrino flux directly, and cannot measure neutrino cross sections without knowing the flux.
This, combined with low statistics, complicated final states and nuclear effects, results in large prior uncertainties on the cross-section model.

\begin{figure}[htbp]
    \centering
    \begin{subfigure}[t]{0.49\textwidth}
        \includegraphics[width=\textwidth]{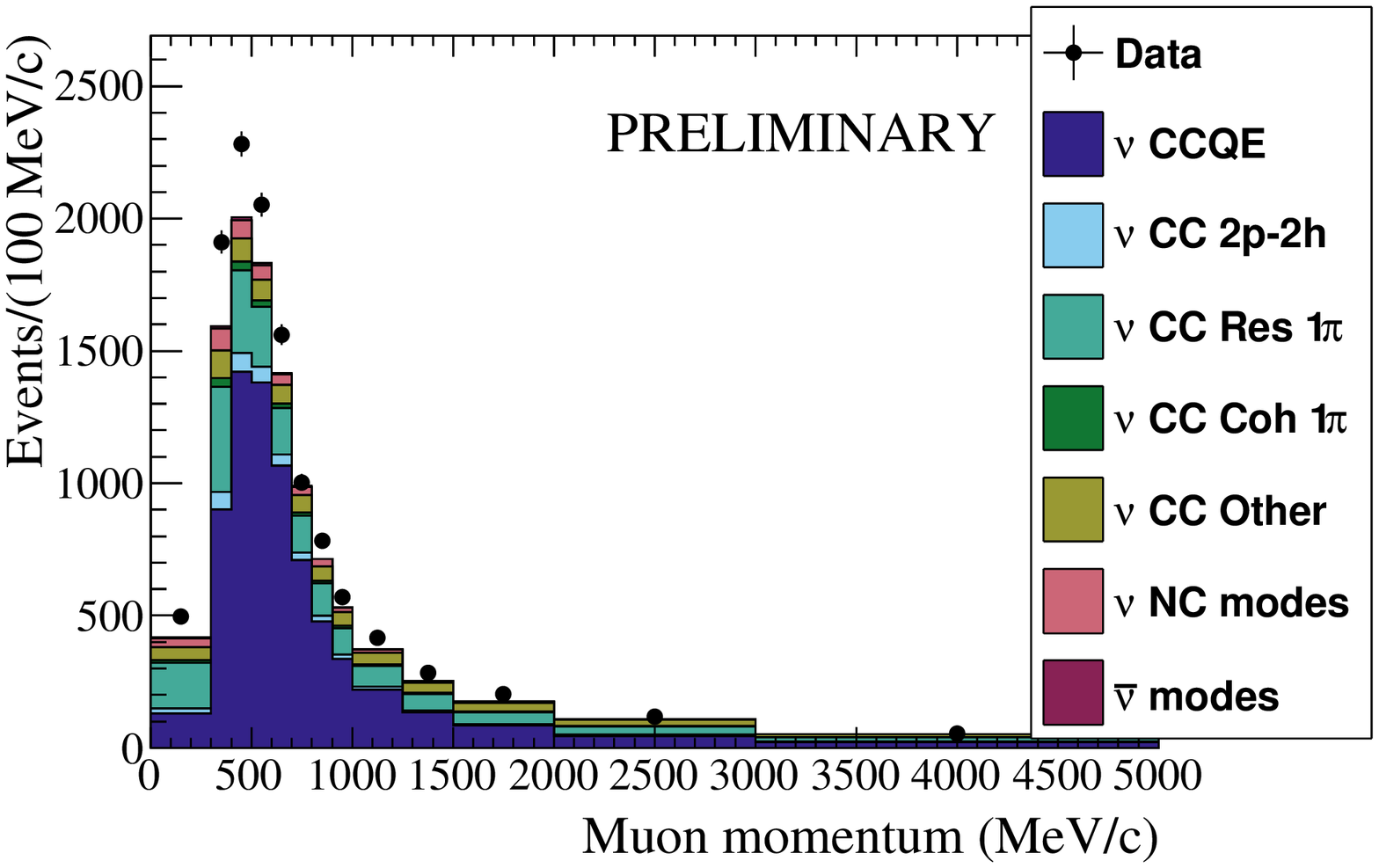}
        \caption{Prefit}
        \label{fig:pre}
    \end{subfigure}
    \hspace{0.001\textwidth}
    \begin{subfigure}[t]{0.49\textwidth}
        \includegraphics[width=\textwidth]{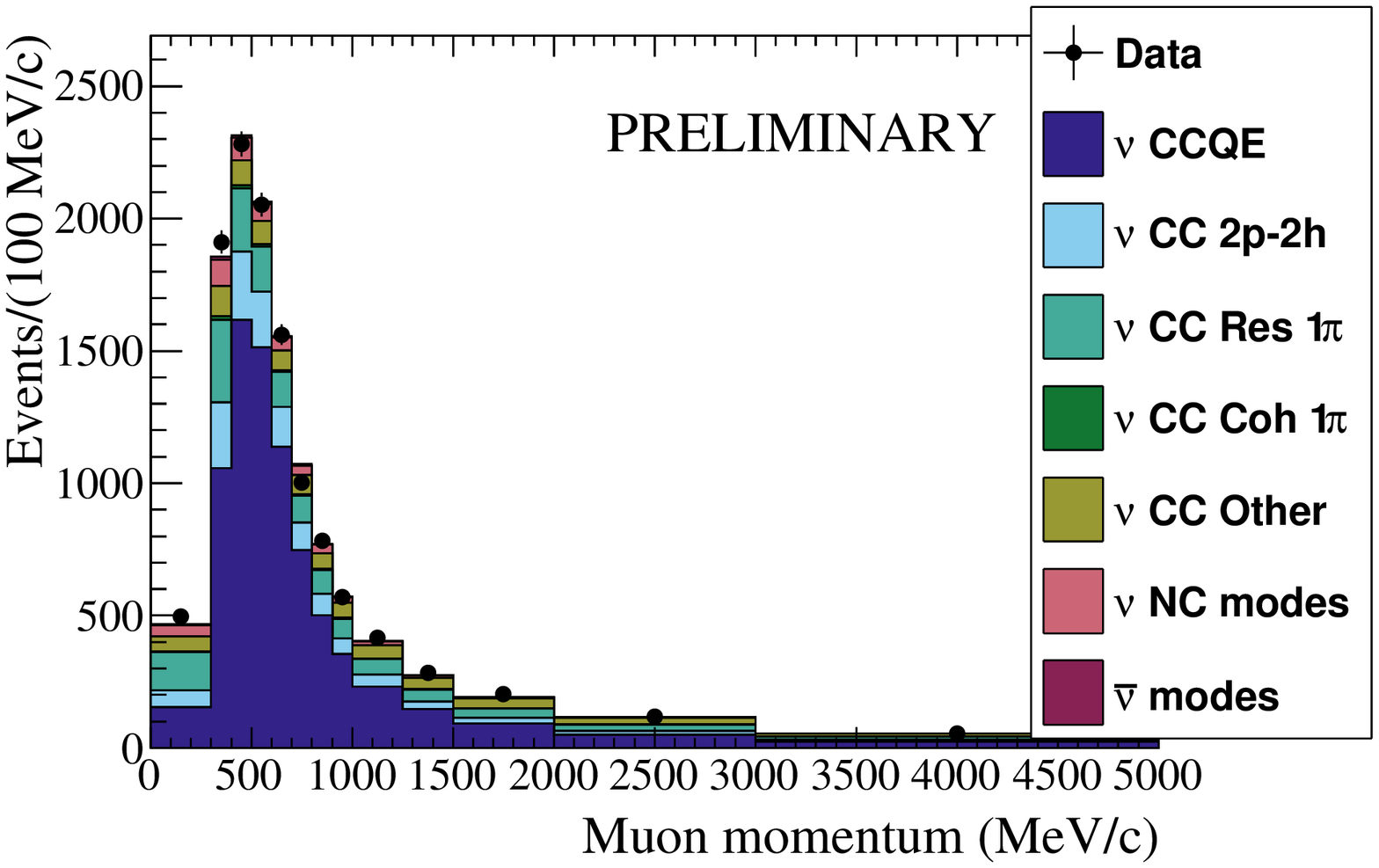}
        \caption{Postfit}
        \label{fig:post}
        \end{subfigure}
    \caption{The ND280 neutrino beam mode $0\pi$ sample before and after the near detector fit.}
    \label{fig:pre_post}
\end{figure}

\begin{figure}[htbp]
    \centering
    \begin{subfigure}[t]{0.55\textwidth}
        \includegraphics[width=\textwidth]{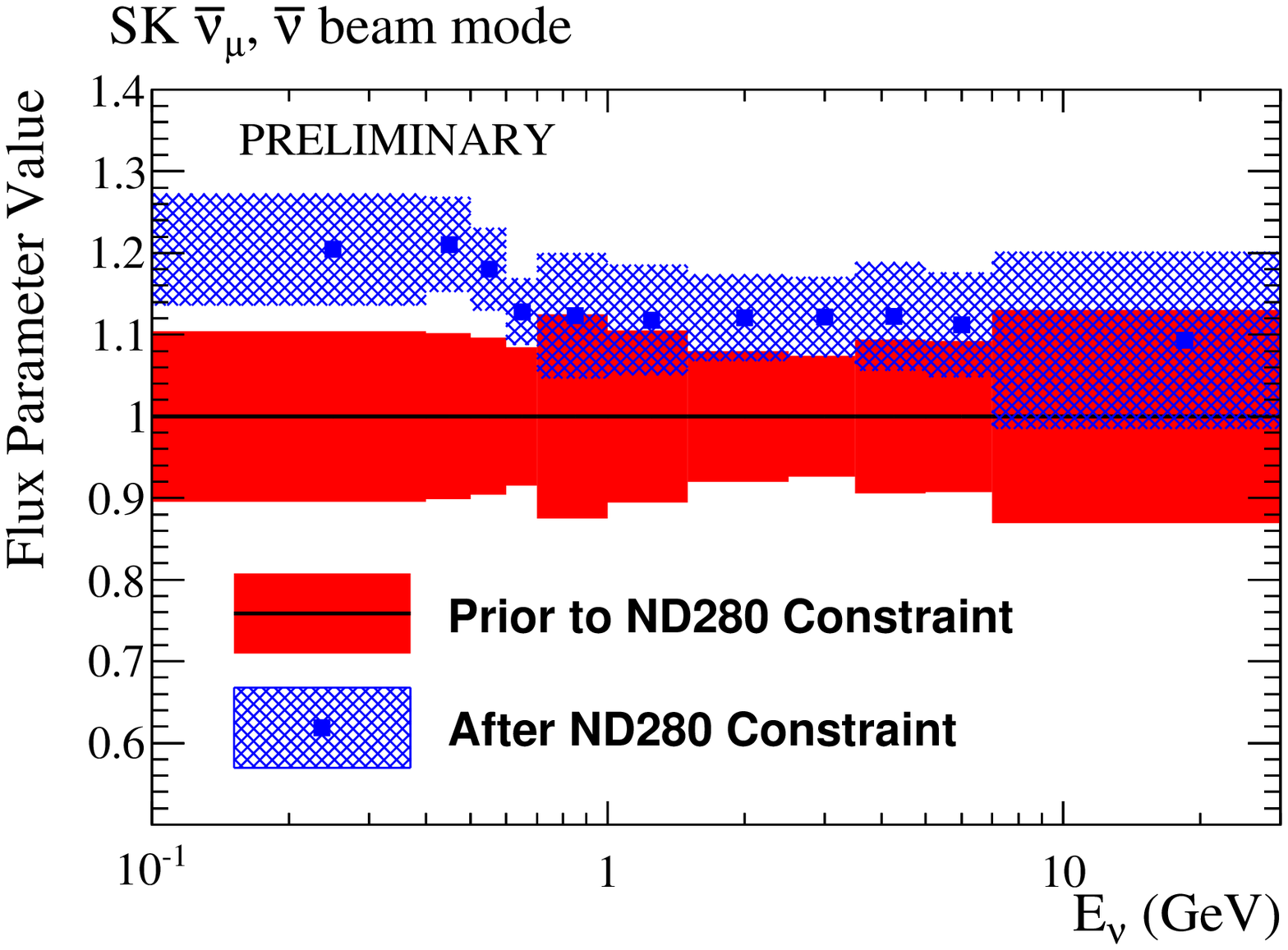}
        \caption{SK anti-neutrino beam mode $\bar{\nu}_{\mu}$ flux parameters}
        \label{fig:sk_flux}
    \end{subfigure}
    \hspace{0.03\textwidth}
    \begin{subfigure}[t]{0.4\textwidth}
        \includegraphics[width=\textwidth]{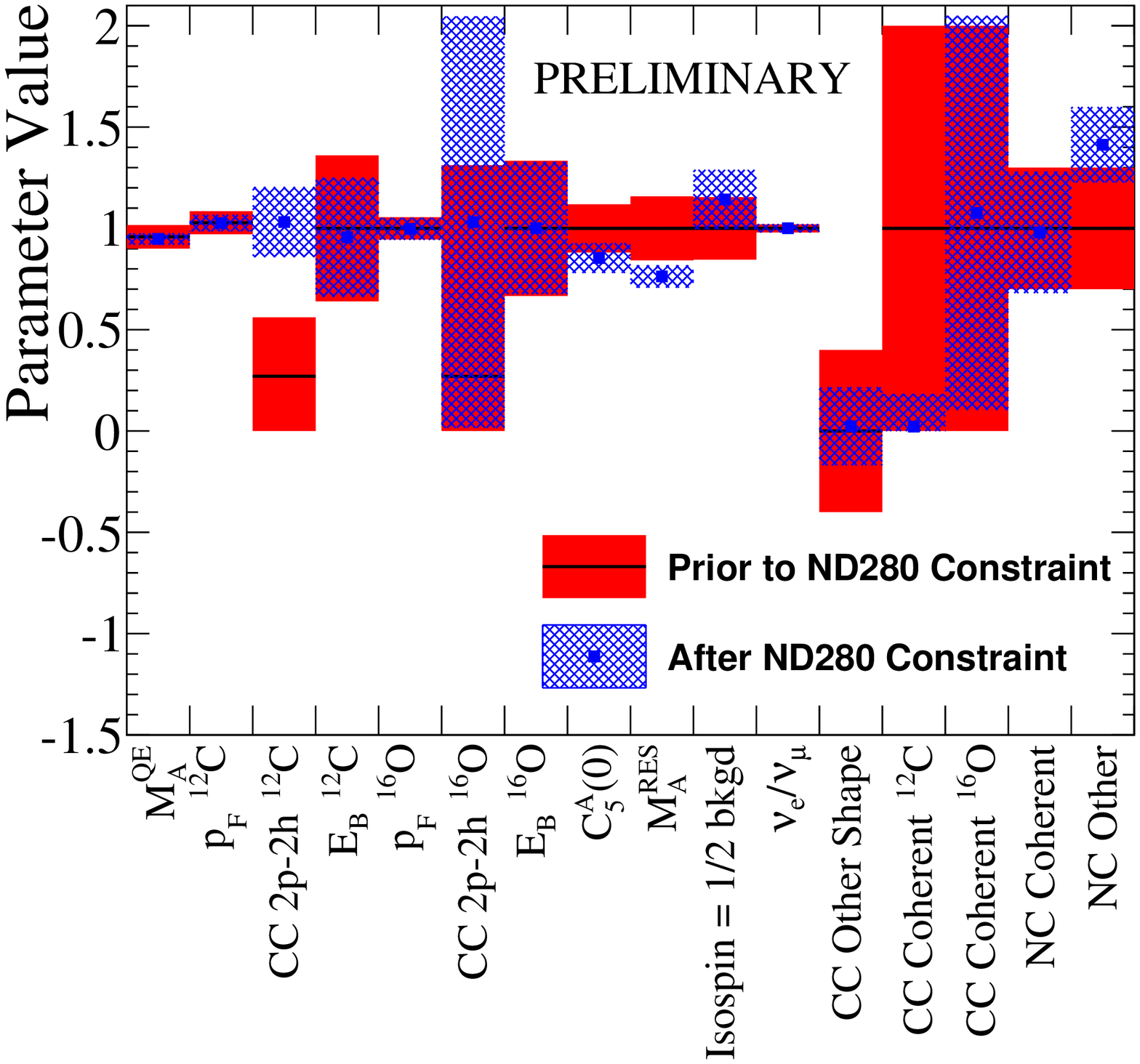}
        \caption{Cross-section parameters}
        \label{fig:xsec}
        \end{subfigure}
    \caption{The fitted flux and cross-section parameter values and uncertainties (blue, hashed region) compared to their priors (red, solid region).}
    \label{fig:fit_params}
\end{figure}

The ND280 analysis performs a simultaneous fit of the 7 event samples to the flux, cross-section and detector models.
Fitting the flux and cross-section togther allows the observed interaction rate at ND280 to constrain the model uncertainties, introducing anti-correlation between the flux and cross-section model parameters.
The neutrino beam mode charged current $0\pi$ sample is presented in Fig.~\ref{fig:pre_post}, showing the improved agreement between data and MC after the ND280 fit.
The fitted flux and cross-section parameter values are shown in Fig.~\ref{fig:fit_params}, with the predicted SK flux as a function of neutrino energy in Fig.~\ref{fig:sk_flux} and the cross-section parameters in Fig.~\ref{fig:xsec}.

\subsection{SK event selection}
\begin{figure}[htbp]
    \centering
    \begin{subfigure}[t]{0.475\textwidth}
        \includegraphics[width=\textwidth]{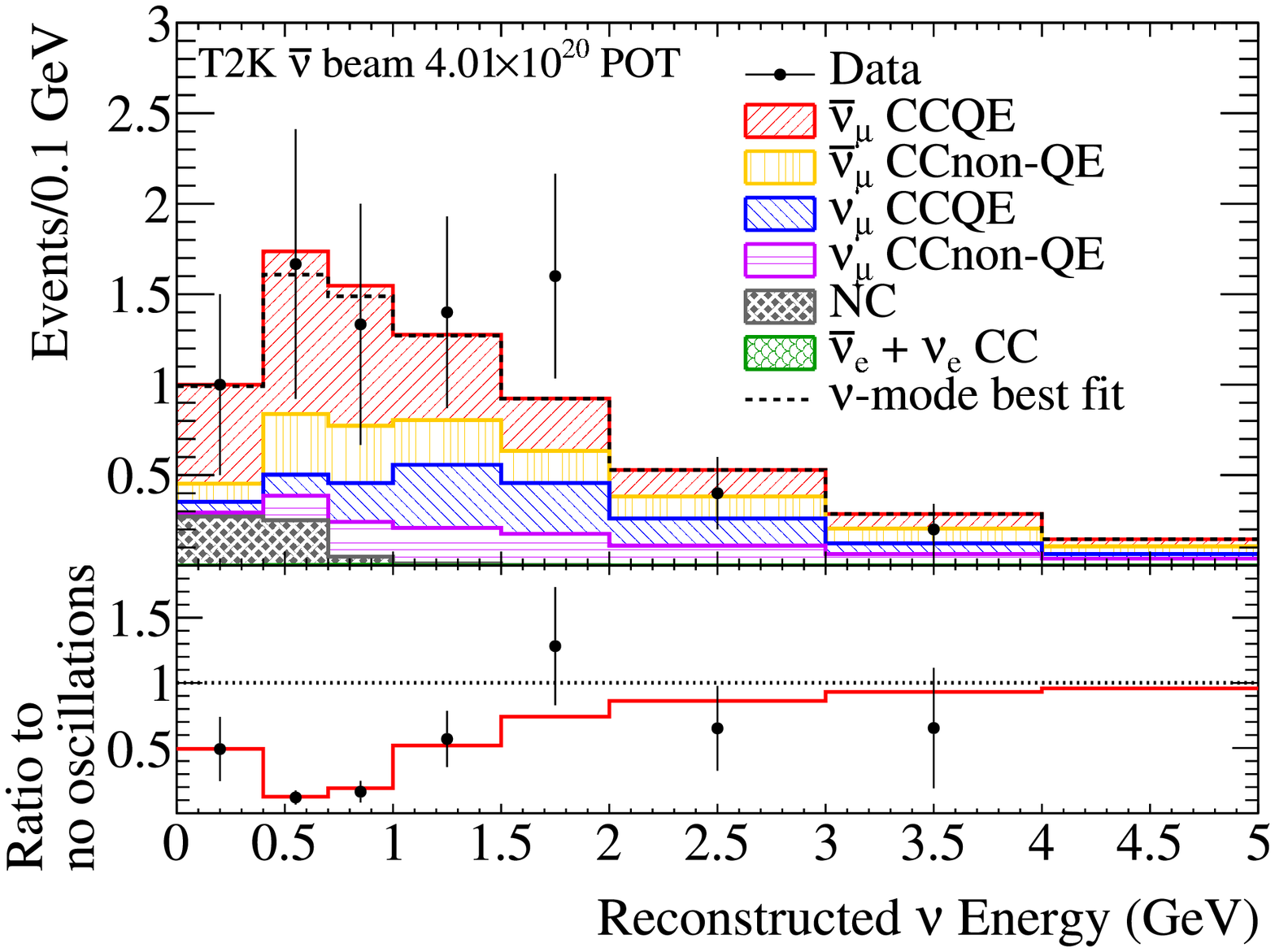}
        \caption{Selected muon-like events compared to the MC prediction~\cite{ref:numub}.  The lower figure shows the ratio of the best fit MC to that predicted with no oscillation.}
        \label{fig:sk_mu}
    \end{subfigure}
    \hspace{0.05\textwidth}
    \begin{subfigure}[t]{0.455\textwidth}
        \includegraphics[width=\textwidth]{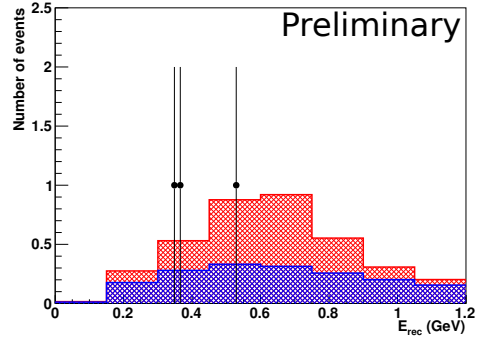}
        \caption{Selected electron-like events, with the expected MC signal (red, light grey) and background (blue, dark grey) overlaid, assuming $\sin^{2}2\theta_{13} = 0.1$ and $\delta_{CP} = -\pi/2$.}
        \label{fig:sk_e}
        \end{subfigure}
    \caption{Selected muon-like~(\ref{fig:sk_mu})~\cite{ref:numub} and electron-like~(\ref{fig:sk_e}) events at SK as a function of reconstructed neutrino energy for $4.01 \times 10^{20}$ POT of anti-neutrino beam mode data.}
    \label{fig:sk_events}
\end{figure}

The SK selection looks for fully contained events with a single Cherenkov ring whose reconstructed vertex is within the SK fiducial volume.
A likelihood is used to determine whether the ring is muon-like or electron-like.
The muon-like candidates are required to have a reconstructed momentum greater than 200~MeV/c and at most one decay electron.
The electron-like candidates are required to have a reconstructed momentum greater than 100~MeV/c, a reconstructed neutrino energy (where the event is assumed to be a charged current, quasi-elastic interaction) below 1250 MeV and no decay electrons.
Electron-like events are passed through a further likelihood discriminator that separates out events containing a $\pi^{0}$.
The resulting event distributions are shown in Fig.~\ref{fig:sk_events}, with the histograms showing the MC prediction with all oscillation parameters set to their best fit values from the T2K neutrino oscillation analysis~\cite{ref:joint} and assuming normal hierarchy.

\subsection{$\bar{\nu}_{\mu}$ disappearance analysis}
As stated earlier, comparisons of $\bar{\nu}_{\mu}$ and $\nu_{\mu}$ disappearance provide a test of the CPT symmetry and can be used to search for non-standard interactions between neutrinos and matter.
The T2K measurement is performed by introducing two additional parameters that control the $\bar{\nu}_{\mu}$ disappearance probability, $\bar{\theta}_{23}$ and $\Delta\bar{m}^{2}_{32}$.
All other oscillation parameters are fixed as shown in Tab.~\ref{tab:oscpars} and the normal hierarchy is assumed.
\begin{table}[!tbp]
    \centering
        \caption{Oscillation parameters used for the fit. The parameters $\sin^2(\overline{\theta}_{23})$ and $\Delta\overline{m}^2_{32}$ were allowed to fit in the ranges given. All other parameters were fixed to the values shown, taken from previous T2K fits~\cite{ref:joint} and the Particle Data Group review~\cite{ref:pdg}.}
        \label{tab:oscpars}
        \begin{tabular*}{0.6\columnwidth}{c@{\extracolsep{\fill}} c@{\extracolsep{\fill}} c}
            \hline
            \hline
            Parameter & $\nu$ & $\overline{\nu}$ \\
            \hline
            $\sin^2(\theta_{23})$ & 0.527 &  fit 0 -- 1  \\
            $\Delta m^2_{32}$ $(10^{-3}$~eV$^2)$ & 2.51 &  fit 0 -- 20 \\
            \hline
            $\sin^2(\theta_{13})$ & \multicolumn{2}{c}{0.0248} \\
            $\sin^2(\theta_{12})$ & \multicolumn{2}{c}{0.304} \\
            $\Delta m^2_{21}$ $(10^{-5}$~eV$^2)$ & \multicolumn{2}{c}{7.53} \\
            $\delta_{CP}$ (rad) & \multicolumn{2}{c}{-1.55} \\
            \hline
            \hline
        \end{tabular*}
\end{table}

A likelihood fit is used to extract $\sin^2(\overline{\theta}_{23})$ and $\Delta\overline{m}^2_{32}$ from the data, marginalising over the systematic uncertainties shown in Tab~\ref{tab:oa_syst}.
The current ND280 analysis does not use data from the water target, so there is a large ND280-unconstrained cross-section error to give a conservative estimate of the uncertainty from extrapolating the results on hydrocarbon at ND280 to water at SK.
\begin{table}[!tbp]
        \caption{Percentage change in the number of 1-ring $\mu$-like events before the oscillation fit from 1$\sigma$ systematic parameter variations, assuming the oscillation parameters listed in Table~\ref{tab:oscpars} and that the anti-neutrino and neutrino oscillation parameters are identical.}
        \label{tab:oa_syst}
        \begin{tabular*}{0.9\columnwidth}{l @{\extracolsep{\fill}} c}
            \hline
            \hline
            Source of uncertainty (number of parameters) & $\delta n^{\textrm{exp}}_{\textrm{SK}}$/$n^{\textrm{exp}}_{\textrm{SK}}$ \\
            \hline
            ND280-unconstrained cross section (6) & 10.0\% \\
            Flux and ND280-constrained cross section (31) & 3.4\% \\
            Super-Kamiokande detector systematics (6) & 3.8\% \\
            Pion FSI and re-interactions (6) & 2.1\% \\
            \hline
            Total (49) & 11.6\%\\
            \hline
            \hline
        \end{tabular*}
\end{table}

\begin{figure}[htbp]
    \centering
    \includegraphics[width=0.5\textwidth]{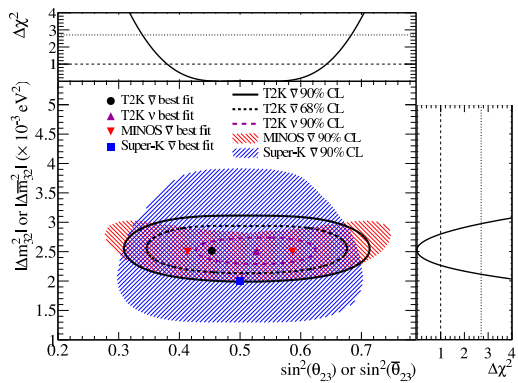}
    \caption{The 68\% and 90\% confidence regions for $\sin^2(\overline{\theta}_{23})$ and $|\Delta\overline{m}_{32}^2|$ assuming normal hierarchy~\cite{ref:numub}. T2K $\nu$~\cite{ref:joint}, MINOS $\overline\nu$~\cite{ref:minos} and SK $\overline\nu$~\cite{ref:sk} 90\% confidence regions are also shown.}
    \label{fig:numubar}
\end{figure}

Fig.~\ref{fig:numubar} shows the results of this fit compared to the T2K neutrino data fit~\cite{ref:joint} and to the anti-neutrino results from the MINOS~\cite{ref:minos} and Super-Kamiokande~\cite{ref:sk} collaborations.
At the best fit point $\sin^2(\overline{\theta}_{23}) = 0.45$ and $|\Delta\overline{m}_{32}^2| = 2.51\times10^{-3}{\mathrm{eV}^2}$, with 68\% confidence intervals of 0.38 -- 0.64 and 2.26 -- 2.80 ($\times10^{-3}$~eV$^2$) respectively.
The results are therefore completely consistent with both previous measurements and the T2K neutrino data, and have been published in Ref.~\cite{ref:numub}.

\subsection{$\bar{\nu}_{e}$ appearance analysis}
Measuring $\bar{\nu}_{e}$ appearance is an integral step before making a measurement of $\delta_{CP}$.
In the PMNS framework the leading term in the $\nu_{e}$ appearance probability depends on both $\theta_{23}$ and $\theta_{13}$, so replacing $\theta_{13}$ with $\bar{\theta}_{13}$ is not sufficient to decouple the neutrino and anti-neutrino oscillation probabilities.
Therefore this analysis assumes that the PMNS matrix describes both neutrino and anti-neutrino oscillation and then introduces a parameter, $\beta$, that independently scales the $\bar{\nu}_{e}$ oscillation probability.

\begin{figure}[htbp]
    \centering
    \includegraphics[width=0.5\textwidth]{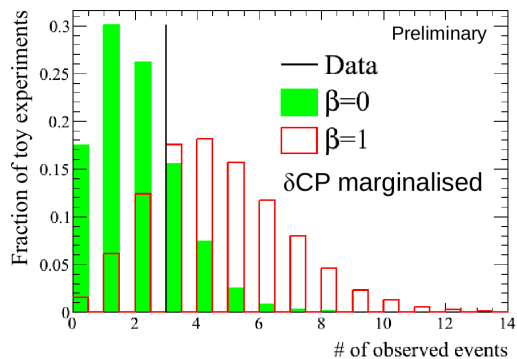}
    \caption{The predicted number of events observed at SK for the $\beta = 0$ and $\beta = 1$ toy experiment ensembles compared to the actual number of observed events in the data.}
    \label{fig:nue_sk_app_figure14}
\end{figure}

For $\beta = 0$ and $\beta = 1$ an ensemble of toy experiments is generated through random throws of the systematics from Tab.~\ref{tab:oa_syst} and all the neutrino oscillation parameters from Tab.~\ref{tab:oscpars} except $\sin^2(\theta_{12})$ and $\Delta m^2_{21}$, shown in Fig.~\ref{fig:nue_sk_app_figure14}.
The $\beta = 0$ ensemble predicts an average of 1.6 electron-like events at SK and the $\beta = 1$ ensemble predicts an average of 4 events, with Fig.~\ref{fig:sk_e} showing the 3 events observed in the data.
This gives a p-value for $\beta = 0$ of 0.26 and a marginalised likelihood ratio for ($\beta = 1$)/($\beta = 0$) of 1.09, showing that the current data does not provide evidence either for or against $\bar{\nu}_{e}$ appearance.

\section{Conclusions}
Having collected $4.0 \times 10^{20}$ POT of data in anti-neutrino beam mode T2K has observed $\bar{\nu}_{\mu}$ disappearance, measuring the $\sin^2(\overline{\theta}_{23}) = 0.45$ and $|\Delta\overline{m}_{32}^2| = 2.51\times10^{-3}{\mathrm{eV}^2}$.
This result is in agreement with the T2K neutrino beam mode data and with existing measurements of $\bar{\nu}_{\mu}$ disappearance.
T2K has also looked for evidence of $\bar{\nu}_{e}$ appearance, observing 3 events compared to a prediction of 1.6 events assuming a $\bar{\nu}_{e}$ appearance probability of zero or 4 events assuming the oscillation probability is equal to that given by the PMNS matrix. 
The current data set does not, therefore, provide evidence either for or against $\bar{\nu}_{e}$ appearance.

T2K is continuing to collect anti-neutrino beam data and working to improve the near detector analysis by including the water target data, culminating in a full joint fit for neutrino and anti-neutrino appearance and disappearance.

\end{document}